\documentstyle [12pt] {article}
\title{SMOOTHED PARTICLE HYDRODYNAMICS CONFRONTS THEORY:
FORMATION OF STANDING SHOCKS
IN ACCRETION DISKS AND WINDS AROUND BLACK HOLES}
\author{Sandip K. Chakrabarti\\
Tata Institute of Fundamental Research, Bombay 400005, India\\
and\\
Diego Molteni\\
Universita' di Palermo, Istituto di Fisica, Via Archirafi, 36\\
90123 Palermo, Italy\\}
\begin{document}
\maketitle
\pagenumbering{arabic}
\begin{abstract}
\baselineskip 24pt
We present results of numerical simulation of
thin accretion disks and winds. We use the Smoothed Particle
Hydrodynamics (SPH) technique for this purpose. We show that the simulation
agrees very well with the recent theoretical work on the shock
formation. The most significant conclusion is that shocks in an inviscid flow
are extremely stable. For the first time, our work also removes the ambiguity
in terms of the location and stability of shocks in adiabatic flows.
\end{abstract}

{\it Subject Headings}:\ {\rm black holes --- stars:accretion ---- stars:winds
--- shocks waves -- hydrodynamics - numerical simulations\\}

\newpage

\baselineskip 24pt

\noindent { 1. INTRODUCTION}

There has been an increasing interest in the study of shock waves
in accretion disks and jets in recent years
(For a recent review, see, Chakrabarti, 1990a). Shock waves
are important for several reasons. Shocks in disks in active
galactic nuclei are thought to contribute to high energy
cosmic rays through {\it in situ} particle acceleration
(Kazanas \& Elison 1986). Standing shocks can contribute to the excess
UV observed in some active galaxies (Chakrabarti \& Wiita 1992).
Non-axisymmetric shocks waves (Sawada, Matsuda \& Hachisu 1986)
provide an extremely efficient transport mechanism of angular momentum in the
disks. Non-stationary shock waves can contribute to the variability in the
spectra of disks around stellar mass (Chakrabarti \& Matsuda 1992)
as well as galactic black holes (Chakrabarti \& Wiita 1993).
Several models of self-consistent shock solutions in spherically
symmetric accretion are also present in the literature (Chang \& Ostriker 1985;
Babul, Ostriker \& Meszaros 1989). Apart from these, shock waves are
invoked in solar winds as well as in jets (Wiita 1992; Hughes,
Aller \& Aller 1985) to explain variabilities.

So far, no numerical simulation has been carried out
which has convincingly proven the
existence of steady shock waves in disks and wind. Extensive theoretical
solution that is present in the literature (Chakrabarti 1990a) provides a
very good understanding of the parameter space spanned by the specific energy
(${\cal E}$) and the angular momentum ($\lambda$ ) for which shock waves
are important. Typically,
if the angular momentum is close to the marginally stable value, and the
initial kinetic energy (for accretion) or thermal energy (for wind)
is within a few percent of the rest mass energy, the flow should
pass through a shock.

One major problem in these studies is that, for a given pair
(${\cal E},\ \lambda$), the shock location is not determined uniquely.
Chakrabarti (1989, 1990a) shows that the question of uniqueness for
adiabatic flows is not resolved by local stability analysis alone.
Secondly, these solutions show that, for the {\it same} initial parameters,
for every solution which includes a shock, there exists another solution
which is shock-free. It was conjectured (Chakrabarti, 1990a),
but not proven, that the solution which {\it includes} a shock must be chosen
since it has more entropy than the shock-free solution. The conjecture
was not tested numerically also. Furthermore, it
remained unresolved, as to which one of the two possible locations of
shock will be chosen by the flow, and whether or not any one of
these solutions are stable. In the present paper, we answer each of these
questions. We use a smoothed particle hydrodynamics code to simulate time
dependent behavior of the infalling matter onto a black hole, or the outgoing
winds from regions close to a black hole. We show that the numerical
results agree with the analytic ones within the limit of numerical
accuracy. We show that an unperturbed flow chooses the shock-free
solution, but as it is perturbed, only one of the two shock
solutions are chosen. The shock is found
to be very stable in the sense that even when the degree of
perturbation is greatly reduced, the shock remains in the flow.

The plan of the paper is the following: in the next Section, we present
one dimensional hydrodynamic equations in cylindrical geometry
and present a few examples of the solution which include Rankine-Hugoniot
shocks. In \S 3, we briefly describe how the SPH code was implemented for
our present case. In \S 4, we present time dependent behavior and
show how, after a transient period, the flow settles down to the theoretical
steady state solution which includes a shock.

\noindent{2. THEORETICAL SHOCK LOCATIONS IN A THIN FLOW AROUND A BLACK HOLE}

We assume a thin, rotating, adiabatic accretion or wind near a black hole.
We take the Newtonian model for the non-rotating central compact object
as given in terms of the Paczy\'{n}ski \& Wiita (1980)
potential. We also assume a polytropic equation of state for the
accreting (or, outflowing) matter, $P=K \rho^{\gamma}$, where,
$P$ and $\rho$ are the isotropic pressure and the matter density
respectively, $\gamma$ is the adiabatic index (assumed in this
paper to be constant throughout the flow, and is related to the
polytropic index $n$ by $\gamma = 1 + 1/n$) and $K$ is related
to the specific entropy of the flow $s$: $s = const$
implies $K = const$. We assume that entropy, and thus $K$ can vary
only at the shock. As there is no dissipation in radial flow, the specific
angular momentum $\lambda$ is constant everywhere. A complete solution of the
stationary model requires the equations of the energy, the angular momentum
and the mass conservation supplied by the transonic conditions at
the critical points and the Rankine-Hugoniot conditions at the
shock. The general procedure followed is the same as is presented
in Chakrabarti (1989, 1990a). However, presently, we
use the model of a strictly thin flow of constant transverse height
(as opposed to a one-and-a-half dimensional flow in Chakrabarti, 1989).
This allows one to test a one dimensional SPH
code. In a later paper (Molteni \& Chakrabarti, 1993), we shall
present the results of the simulation of a fully two dimensional SPH code.

\noindent{2.1 The Model Equations}

In what follows, we use the mass of the black hole $M$, the velocity of
light $c$ and the Schwarzschild radius $R_g=2GM/c^2$ as the units of
mass, velocity and distance respectively. The dimensionless energy
conservation law can be written as,
$$
{\cal E} = \frac{\vartheta^2}{2}+\frac{a^2}{\gamma-1}
+ \frac{\lambda ^2}{2x^2}+g(x)
\eqno{(1a)}
$$
Here $g(x)$ is the radial force potential, which in the pseudo-Newtonian
model takes the form: $g(x) = - {1}/{2} (x-1)^{-1} $.
Here, $\vartheta$ and $a$ are the non-dimensional radial and the sound
velocities, $x$ is non-dimensional radial distance. Apart from an unimportant
geometric factor, the mass conservation equation is given by,
$$
{\dot M} = \vartheta \rho x h_0
\eqno{(b)}
$$
where $h_0$ is the constant half-thickness of the flow.
It is useful to write the mass conservation equation in terms of
$\vartheta$ and $a$ in the following way,
$$
\dot{\cal M} = \vartheta x a^{2n}
\eqno{(2)}
$$
We shall use the word
`accretion rate' for this quantity, keeping in mind, however, that
${\dot {\cal M}}\sim {\dot M}{K^n}$ does not
remain constant at the shock because of the generation of entropy.
The shock conditions which
we employ here are the following (subscripts `-' and `+' refer to
quantities before and after the shock): The energy conservation equation,
$$
{\cal E}_+ = {\cal E_-},
\eqno{(3a)}
$$
the pressure balance condition,
$$
P_+ + \rho_+ \vartheta_+^2 = P_- + \rho_- \vartheta_-^2
\eqno{(3b)}
$$
and the baryon number conservation equation,
$$
\dot M_+ = \dot M_-
\eqno{(3c)}
$$
In order to have a shock, the flow must be supersonic,
i.e., the stationary flow must pass through a sonic
point. The sonic point conditions are derived following the
general procedure (Chakrabarti 1990a) by first
differentiating the energy and mass conservation equations (eqns. 1a and 2)
and eliminating ${da}/{dx}$ from them to obtain,
$$
\frac{d\vartheta}{dx} [\vartheta - \frac {a^2}{\vartheta}] =
\frac{a^2}{x} - \frac{dG}{dx}
\eqno{(4)}
$$
where, $G(x)={\lambda^2}/{2x^2} +g(x)$ is the effective
potential. The left-hand side and the right-hand side vanish
simultaneously at the critical points in order that the stationary solution
is smooth everywhere. The vanishing of the left-hand side requires,
$$
v_c(x_c)=  a_c(x_c).
\eqno{(5a)}
$$
The vanishing of the right-hand side gives the sound speed
at the outer critical point as,
$$
a_c^2(x_c)=\frac{\lambda_K^2-\lambda^2}{x_c^2}
\eqno{(5b)}
$$
The subscript $c$ denotes quantities at the critical points.
Eqns. (1a), (2), (3a-c), (5ab) are simultaneously solved
to obtain the full set of solutions which may include shocks.
The general procedure is already discussed in detail in Chakrabarti
(1989, 1990a) and will not be repeated here. We only present a few
examples of the solutions below.

\noindent{2.2 Examples of Shock Locations}

Physically, a shock can form when the rotating flow is close to the centrifugal
barrier. However, because the flow has significant pressure,
it is not essential that the angular momentum should
be so high as to have an actual barrier. That is, flow angular momentum
could be much lower than the marginally stable value ($\sim 1.83$ in our unit).
We provide here two examples, one with a high angular momentum and the
other with a low angular momentum.
Fig. 1(a-b) shows the contours of constant ${\dot {\cal M}}$, which,
for a given mass accretion rate, measures the specific entropy of the
flow. In (a) ${\cal E}=0.01095,\ \lambda=1.90$ and in (b) ${\cal
E}=0.0557, \ \lambda=1.75$ are used as the input parameters. The arrowed
curves show the stationary solution for accretion and the vertical lines
show the possible shock transitions. In (a), we retain
the notation of Chakrabarti (1989, 1990a) regarding the shock locations.
Thus, the shock located at $X_{s2}$ is closer to the black hole than
the shock at $X_{s3}$. (The other two locations $X_{s1}$ and $X_{s4}$
are irrelevant in the present context.) An accreting flow, subsonic at a large
distance first passes through the outer sonic point at $O$, then passes through
a shock, located either at $X_{s2}$ or at $X_{s3}$ and subsequently, falls
onto the black hole after going through the inner sonic point $I$. The
flow passing through inner sonic point has a considerably higher
entropy than the flow passing through the outer sonic point.
Thus, for example,
in Fig. 1a, ${\dot {\cal M}_-}=0.62\times 10^{-6}$, ${\dot {\cal M}_+}=
0.105\times 10^{-5}$. In Fig. 1b, ${\dot {\cal M}_-}=3.295\times 10^{-4}$,
${\dot {\cal M}_+}= 3.422\times 10^{-4}$.
Theoretically, both the locations satisfy all the shock conditions.
In an wind, the situation is exactly the opposite. There the entropy
of the flow passing through the outer critical point is higher
than that passing through the inner sonic point.
Fig. 2(a-b) show the location of the shocks as a function of the specific
energy of the flow. One can obtain similar variations for wind solutions
also (see, Chakrabarti 1989, 1990a).

Chakrabarti (1992), using a simple consideration of varying (total)
pressure at the post-shock flow as the shock is perturbed,
shows that for adiabatic flow, $X_{s3}$ for winds is definitely unstable and
$X_{s3}$ for accretion is definitely stable. Thus, it became
clear that $X_{s2}$ for accretion can not be stable.
This argument cannot, however, determine the fate of $X_{s2}$ for
the winds positively. We present the argument below.

Using the definition of accretion rate (1a) and the net pressure (3a)
presented above, we obtain the variation of the total pressure of the flow as,
$$
\frac{1}{\dot M} \frac{d\Pi}{dx}=
\frac{1}{\dot M} \frac{d (P+\rho \vartheta^2)}{dx}
= \frac{1}{\vartheta x}[\frac{\lambda^2}{x^3}-
\frac{1}{2(x-1)^2}-\frac{\vartheta^2}{x}]
$$
or, using equation (4),
$$
\frac{1}{\dot M} \frac{d\Pi}{dx}=- \frac{(a^2-\vartheta^2)}{\vartheta^2 x}
\frac{d\vartheta}{dx}-\frac{(a^2+\vartheta^2)}{\vartheta x^2}
\eqno{(6)}
$$
Since the second term is negative definite, the pressure variation,
and hence the stability property
depends strongly on the sign of the derivative $d\vartheta/dx$. If, for
example, after the perturbation of the shock location, the pressure on the
post-shock flow becomes higher than the equilibrium solution, the shock
returns back to the original location and the shock is stable. In this context,
we wish to point out that the perturbation of the post-shock region only
matters, since the pre-shock flow is supersonic. In the accretion flow or
winds, $d\vartheta/dx$
in the post-shock branch is negative at $X_{s2}$ and positive at $X_{s3}$.
Remembering, that the flow is along {\it decreasing} $x$ for accretion, and
along {\it increasing} $x$ for winds, the above behavior of $d\vartheta/dx$
implies that, for accretion, $X_{s3}$ is definitely stable, and hence
$X_{s2}$ is definitely unstable. Similarly, for winds, $X_{s3}$ is definitely
unstable, but the fate of $X_{s2}$ can not be ascertained by the above
argument. We have studied the stability of shock waves
in adiabatic flows in this paper. recent global stability analysis
indicates (Nakayama, 1992) that $X_{s2}$ may be unstable in the
{\it isothermal} accretion also.

\noindent {3. NUMERICAL SIMULATION OF ACCRETION DISKS AND WINDS}

The Smoothed Particle Hydrodynamics (SPH) method that we use
has been primarily
developed to deal with the fluid dynamics in astrophysical context
(Lucy 1977; Gingold \& Monaghan 1977; for a recent review, see, Monaghan,
1992). It is an interpolating method, i.e. the fluid properties
are locally approximated in terms of its values at a set of disordered
points (the ``pseudo-particles"). It has many
attractive features that make its use particularly appealing: (a)
its Lagrangean character permits easy description of
fluid motion; (b) it does not require a grid which allows
computer memory saving, specially for 3-D problems with large
voids in the integration domain; (c) it conserves energy, linear momentum and
angular momentum quite accurately; (d) it is easy to implement,
the difficulty involves a search of neighboring particles,
the interactions of which are to be taken into account.
For a detailed description of the method see Monaghan, (1985, 1992).
Molteni and Giannone (1992) has recently implemented the code
in cylindrical coordinates for problems with axial symmetry.
The basic point for the cylindric geometry approach is to assume
the interpolating Kernel $W$ which is a function of cylindrical radial
coordinate ${\vec r}={\vec r}(x,z)$ and $k$-th particle of mass $m_k$ as,
$$
m_k=2\pi\rho_k \Delta {\vec r}_k
\eqno{(7)}
$$
Any smooth function $A({\vec r}_i)$ at ${\vec r}_i$ is defined as,
$$
A({\vec r}_i)=\int A({\vec r}) W({\vec r}-{\vec r}_i;h)\frac{2\pi\rho x}
{2\pi\rho x} d{\vec r} \approx
\sum_k \frac{m_k}{2 \pi x_k \rho_k} A({\vec r}_k) W({\vec r}_k-{\vec r}_i; h),
\eqno{(8)}
$$
$h$ being the particle size.
Thus, for example, for the density at each point, we have the simple
expression that identically satisfy the continuity equation in the cylindrical
coordinate:
$$
\rho({\vec r}_i)\approx\sum_k\frac{m_k}{x_k} W({\vec r}_k-{\vec r}_i;h)
\eqno{(9)}
$$

The equations of motion to be solved consist of the radial momentum equation:
$$
\left(\frac{Dv_x}{Dt}\right)_i=-\sum_k\frac{m_k}{x_k}
\left(\frac{P_i}{\rho_i^2}+\frac{P_k}{\rho_k^2}+\Pi_{ij}\right) \frac{\partial
W_{ik}}{\partial x_i}+\frac{\lambda^2}{x_i^3}
-\frac{1}{2(r_i-1)^2}\frac{x_i}{r_i} ,
\eqno{(10a)}
$$
the vertical component of the momentum equation
$$
\left(\frac{Dv_z}{Dt}\right)_i=-\sum_k\frac{m_k}{x_k}
\left(\frac{P_i}{\rho_i^2}+\frac{P_k}{\rho_k^2}+\Pi_{ij}\right) \frac{\partial
W_{ik}}{\partial z_i} -\frac{1}{2(r_i-1)^2}\frac{z_i}{r_i}
\eqno{(10b)}
$$
and the energy equation
$$
\left(\frac{D{\cal E}}{Dt}\right)_i=-\frac{1}{2}\sum_k\frac{m_k}{x_k}
\left(\frac{P_i}{\rho_i^2}+\frac{P_k}{\rho_k^2}+\Pi_{ij}\right)
\frac{x_iv_i-x_kv_k}{x_i}\nabla_i W_{ik}
\eqno{(10c)}
$$

Here,
$$
\Pi_{ij}=\frac{\alpha \mu_{ij} {\bar c}_{ij} +\beta \mu_{ij}^2}
{{\bar \rho}_{ij}}
$$
$$
{\bar c}_{ij}=\frac{c_i+c_j}{2}
$$
$$
{{\bar \rho}_{ij}}=\frac{\rho_i+\rho_j}{2}
$$
$$
\mu_{ij}=\frac{x_iv_x_i-x_jv_x_j}{x_i(l_{ij}^2+\eta^2)}+
\frac{(v_z_i - v_z_j)(z_i-z_j)}{(l_{ij}^2+\eta^2)}
$$
$$
l_{ij}^2=(x_i - x_j)^2 + (z_i - z_j)^2
$$
$$
\eta=0.1 h^2
$$
$\alpha$ and $\beta$ are artificial the viscosity coefficients
used to damp out oscillations in shock transitions.

This implementation of the SPH in cylindrical code is similar in
every respect to the original SPH code in Cartesian coordinate,
except that the mass of a particle appears divided by its axial distance
and the relative velocity $v_k-v_i$ between $k$-th and $i$-th particles
must be replaced by a $(x_kv_k-x_iv_i)/x_i$ term.

\noindent {4. NUMERICAL RESULTS}

The results described in this Section are strictly for
thin flows, i.e., the flow is of constant vertical height
and it has no vertical structure. The inner boundary is kept at
$x=1.1$ and the absorbing boundary condition is used to mimic the
black hole properties. (Note that the black hole horizon is at $x=1$
in our unit.) We use the particle size $h$,
which is of importance in SPH method, in a manner so that the result
is reasonably accurate where shocks are expected to form. An arbitrarily
small $h$ would increase accuracy, but the number of particles injected
would have been extremely high, increasing the computing time.
Typically, for an accurate result at a distance
$x$, one needs to consider $h\leq \mid G(x)/(dG(x)/dx) \mid$, $G(x)$
being the effective potential given in \S 2. The criterion for particle
injection into the disk is determined by the size of the injection sphere.
As the size of the injection region $S_i$
gets smaller compared to the particle size, the injected velocities in the
disk have lesser dispersion, or perturbation. We have tested the code
for a wide range of these parameters. The characteristics of the shocks,
if formed, are not sensitive to these parameters. Generally speaking,
if the values of particle size $h$ and $S_i$ are very small, the flow is
perturbation free, and as expected, no shock is formed. As the size $S_i$ is
made bigger, the flow starts with a larger dispersion of velocities
at the injection point. The injection point of matter
is kept at the outer edge of the disk for the
study of accretion process and at a point close to the hole for the
study of wind propagation.

We first carry out the simulation of accretion process corresponding
to the cases shown in Figures 1a and 2a. In Fig. 1a, we observe that
if the flow is initially unperturbed, it takes the route $OU$, otherwise,
it passes through a shock as the arrowed curve shows.
Fig. 3a shows the results of simulation. Matter is injected at $X=30.4$
with the sound speed and velocity according to the analytical values at that
point. The dashed curves are the
analytical variation of Mach number with distance. For a choice
$h=0.2$, and $S_i=0.05$, we obtain the solution (not shown)
lying practically on the supersonic branch of the
analytical result. As the injection region size is raised, the shock
appears very quickly. The successive solid curves show how the
shock travels backwards with time
and eventually, remains stable practically at the
place where theoretically $X_{s3}$ (at $23.5$) is predicted (shown
in vertical dashed line). After this steady shock formation,
we reduced the size $S_i$  again to $0.05$ to check if the
shock goes away. We find that after an initial inward drift,
the shock goes back again to the location of $X_{s3}$ and remains stable there.
Notice that the pre-shock and the post-shock solutions of the simulation
are effectively on the theoretical curves. In the simulation,
we chose $\alpha=\beta=1$ for the artifical
viscosity parameters to smooth out the oscillations at both sides
of the shock (though a little oscillation is still visible).
With a higher viscosity, the shock becomes diffusive and spreaded out.

In another example, in order to show that shocks (albeit weaker) can form even
when the angular momentum of the flow is much less than the marginally
stable value (i.e., even when there is no Keplerian orbit and no
centrifugal barrier) we present the simulation when
$\lambda=1.75$. The corresponding theoretical results are shown
in Figures 1b and 2b. We show the results for the same $h$ and $S_i$
parameters as in the previous paragraph.
The shock is formed and remained stable at the same place as $X_{s3}=
6.04$ (shown in dashed line). We first use $\alpha=1$ and $\beta=1$,
but the shock was diffusive since it is very weak
(the curve marked T=7000). We then reduced
$\alpha$ and $\beta$ to $0.5$. The shock becomes steeper, though with a small
oscillation. Flow was injected at $X=14$ in this case.

In the case of winds, matter was injected closer to the hole. At $X=100$
the matter is removed from the region of integration. In the
particular example, the time evolution of which
is shown in Fig. 4, we chose $\lambda=1.90$ and ${\cal E}=0.01725$. Matter
was injected at $X=2.3$. The shock is first formed at a larger distance.
It traveled backward, and remained stable at the location of the
shock $X_{s2}=2.7$ predicted theoretically (shown by vertical dashed line)
for this case.

In both the examples of shock formation in accretion flows provided above, the
first occurrence of the shock took place close the inner sonic point $I$.
Subsequently, it moved away towards $X_{s3}$ very rapidly.
The shock predicted at $X_{s2}$ was not formed at all. We have observed
this behavior in all the cases we studied. Thus, it may be assumed that the
inner shocks are completely unstable and only $X_{s3}$ survives,
exactly as discussed in \S 2.2.
In the case of the wind solution that we presented,
the shock predicted at $X_{s3}$ is not formed. Initially,
the shock is formed close to the outer sonic point, but it rapidly
moved towards $X_{s2}$. In simulations of wind flows, wherever shocks
formed, only the inner shock at $X_{s2}$ is found to be stable. Sometimes,
however, even though shocks have been predicted theoretically, no stable
shock is found to be present. Such results occur when the entropy
jump between two transonic solutions (one passing through $O$ and
the other passing through $I$) is so high that the strength of the shock
is very large (about $5$ or more). In such cases,
the perturbation on the supersonic branch does not seem to grow
sufficiently to reach the corresponding subsonic branch. Eventually,
the perturbation dies out and the flow follows the shock-free solution.
Thus, we believe that no strong shock may be formed in wind flows.
This behavior is fully consistent with the discussion in \S 2.2, where it was
predicted that ${X_{s2}}$ for winds may or may not be stable.
In passing, we may remark that we have done several simulations
where, theoretically no shock is expected. We do not find any
shock formation in the numerical simulations either. Even when large
perturbations are introduced, the flow seems to return back to its
steady supersonic solution after a short period of transient state.
In the literature, there have been quite a few suggestions regarding the nature
of the discontinuous transitions in transonic flows. Following Liang
\& Thompson (1980), who first observed the multiplicity of saddle type sonic
points in flows which include angular momentum (which is so vital
in having shock transition in the disk), Abramowicz \& Zurek (1981)
suggested the possible occurance of "bistability" in the flow as it may
sometimes choose a disklike and sometimes Bondi-like solution.
In our simulation it is seen that the flow
becomes {\it steady} asymptotically and smoothly passes through
{\it both} the saddle type sonic points simultaneously without
ever shuttling back and forth and causing bi-periodicity. Bi-stability is not
expected either, since, for a given ${\cal E}$ and $\lambda$, the flow will
have to have different entropies at two saddle points
(Chakrabarti 1989). Subsequently, Wallinder, Kato \& Abramowicz (1992)
suggested a discontinuity which involves transitions between
certain topologies (see Fig. 9b of their paper). None of our
simulations show any such transitions, because, theoretically,
the entropy of the flow has to {\it decrease} in order to have
such a discontinuity. Detailed properties of the time-dependent
behavior and the variability of the emitted spectra that it introduces
will be presented elsewhere.

Detailed properties of the time dependent behavior and the variability of
the emitted spectra that it introduces will be presented else where.

\noindent {5. CONCLUDING REMARKS}

In this paper, we have studied the nature of the Rankine-Hugoniot
shocks in thin, axisymmetric accretion and winds near a compact object
both theoretically and through numerical simulation.
The agreement in these two approaches is found to be excellent.
Our numerical simulation indicates that shocks could be common in
accretion disks and winds. One of the most significant results of our
simulation is that, for the first time, it is realised that the
flow chooses only one of the two possible shock locations,
and does not, for example, shuttle between one to the other. In other
words, no behavior resembling some kind of `bi-periodicity' is seen. Similarly,
for the time, our result shows that given two choices, a shock-free
solution and a solution which includes a shock, the flow chooses
the shock-free solution only if it is truly unperturbed. Any significant
perturbation quickly introduces a shock. The shock
does not disappear even after the perturbation is gradually reduced.
This suggests that a realistic astrophysical disk is likely to have
shocks in them. In a two-Dimensional simulation (Molteni
and Chakrabarti, 1993) it is observed that one need not introduce any external
perturbation at all, since the flow inherently contains turbulence.

So far, we have not studied the fate of the shock formation in viscous flows.
In the case of isothermal flows, analytical work (Chakrabarti, 1990b) indicates
that when viscosity is sufficiently high only one of the solutions, namely,
$X_{s2}$ for accretion and $X_{s3}$ for winds remains. Our present
experience suggests that since $X_{s2}$ is unstable for accretion and
$X_{s3}$ is unstable for winds, when the viscosity is sufficiently
large, the flow should not have a shock at all. For a smaller viscosity,
however, one should still expect a weak shock formation (Chakrabarti 1990b)
in the flow.

The result of the simulations presented here agrees so remarkably well
with the analytical results that we are confident that the
code can be used for a complete two dimensional disk models,
for which no complete theory of shock formation is present. Our preliminary
runs indicate that strong shocks form even when the disk is truly `thick',
except that the shock becomes weaker with height. The locations of the
shocks also correspond to roughly the locations predicted by the
1.5D model (Chakrabarti 1989). Details of these
simulations will be published else where  (Molteni and Chakrabarti, 1993).

\newpage

\centerline {\large REFERENCES}

\noindent Abramowicz, M.A., \& Zurek, W. 1981, ApJ, 246, 314\\
Babul, A., Ostriker, J.P., \& Meszaros, P., 1989, ApJ, 347, 59\\
Chakrabarti, S.K., 1989, ApJ, 347, 365\\
Chakrabarti, S.K., {\it Theory of Transonic Astrophysical Flows}, 1990a,
World Scientific Publ. Co. (Singapore)\\
Chakrabarti, S.K., 1990b MNRAS, 243, 610\\
Chakrabarti, S.K., 1992, Presented at the 13th International Conference
on General Relativity and Gravitation, Cordoba, Argentina (28th June
- 4th July)\\
Chakrabarti, S.K, \& Wiita, P.J. 1992, ApJ, 387, L21\\
Chakrabarti, S.K., \& Wiita, P.J. 1993, ApJ, (submitted)\\
Chakrabarti, S.K., \& Matsuda, T., 1992, ApJ, (May 10th issue)\\
Chang, K.M., \& Ostriker, J.P. 1985, ApJ, 288, 428\\
Gingold, R.A. \& Monaghan, J.J., MNRAS, 1977, 181, 375\\
Hughes, P.A., Aller, H.D., \& Aller, M.F. 1985, ApJ, 298, 301\\
Hawley, J.W., Smarr, L. \& Wilson, J. 1985, ApJ, 277, 296\\
Kazanas, D. \& Ellison, D.C. 1986, ApJ, 304, 178\\
Liang, E.P.T. \& Thompson, K.A. 1980, ApJ, 240, 271\\
Lucy L.,  Astron. J., 1977, 82, 1013\\
Molteni, D. \& Chakrabarti, S.K., 1993, (in preparation)\\
Molteni, D. \& Giannone, G., J. Comp. Phys., 1992 (submitted)\\
Monaghan J.J., Comp. Phys. Repts., 1985, 3, 71\\
Monaghan J.J., Ann. Rev. Astron. Astrophys., 1992, 30, 543\\
Nakayama, K. 1992, MNRAS, 259, 259\\
Paczy\'{n}ski, B. \& Wiita, P.J. 1980, A\&A, 88, 23\\
Sawada, K., Matsuda, T. \& Hachisu, I. 1986, MNRAS, 219, 75\\
Wallinder, F., Abramowicz, M.A. \& Kato, S., 1992, AA Rev, 4, 79\\
Wiita, P.J., 1990, in Beems \& Jets in Astrophysics,
Ed. Hughes, P.A. (Cambridge University Press), 379 \\

\newpage

{\centerline {\bf FIGURE CAPTIONS}}

Fig. 1ab: The phase space diagram of the accretion flow for the
parameters (a) ${\cal E}=0.01095, \ \lambda=1.90$,
and (b) ${\cal E}=0.05571, \ \lambda=1.75$.
Along the abscissa is the logarithmic radial distance $X$ and along the
ordinate is the Mach number of the flow.
The contours are of constant accretion rate ${\dot{\cal M}}$.
The relevant shock locations are: $X_{s2}=2.6$, and
$X_{s3}=23.6$ in (a) and $X_{s2}=4.4$, and
$X_{s3}=6.04$ in (b) respectively. The inner and the outer critical
points are denoted by $I$ and $O$ respectively.  Theoretical
shock transitions are shown by arrows.

Fig. 2ab:
Variation of shock locations in accretion (solid) and winds
(dashed) as a function of specific energy
${\cal E}$ and angular momentum (a) $\lambda=1.90$ and (b) $\lambda=1.75$.

Fig. 3ab: The results of numerical simulation for accretion flows,
with parameters identical to the cases drawn in Figs. 1a and 1b
respectively. The dashed curves show analytical pre-shock and post-shock
branches. Simulated flow, initially unperturbed,
lie on the supersonic branch. When the flow is perturbed shocks
are produced, which evolved until they reach the site of the analytical
location at $X_{s3}$ (solid curves). Some intermediate evolutionary phases
are marked with time since the beginning of the simulation. Vertical
dashed lines show theoretical shock location. In (b), final shock locations
with higher (marked with T=7000) and lower viscosities are shown for
comparison.

Fig. 4: Example of a numerical simulation of shock formation in winds
close to a black hole. The parameters are: $\lambda=1.90$
and ${\cal E}=0.01725$. The dashed curves correspond to the pre-shock
and post-shock branches computed analytically, and the dashed vertical line
at $X_{s2}=2.7$ is the analytical shock location. The solid curves
(marked with time since the beginning of the simulation)
indicate how matter evolves with time initially and then settles
down at the solution with shock.

\end{document}